\documentstyle[12pt,frascatiphys,rotating,epsfig]{article}
\makeatletter
\def\anti#1{\mathchoice{\@anti{\displaystyle}{#1}}
                       {\@anti{\textstyle}{#1}}%
                       {\@anti{\scriptstyle}{#1}}%
                       {\@anti{\scriptscriptstyle}{#1}}#1}
\def\@anti#1#2{\setbox0\vbox{$#1#2$}
  \rlap{$#1\kern0.25\ht0\overline{\kern-0.25\ht0\phantom{#2}}$\hss}}
\makeatother
\def\spig{\kern.5ex\raise1.2ex\hbox{$|$}\kern-.48em\to}
\begin{document}
\title{ ANALYSIS OF SUBSTRUCTURES IN CHARM DECAYS }
\author{
Sandra Malvezzi \\
{\em Istituto Nazionale di Fisica Nucleare--Sezione di Milano} \\
{\em Via Celoria 16, I-20133 Milan, Italy} \\ }
\maketitle
\baselineskip=14.5pt
\begin{abstract}
Recent results on Dalitz analysis of three-pseudoscalar decays are 
discussed in the context of probing charm hadronic-decay mechanisms: 
the role of FSI effects, which create phase shifts between the interfering 
resonant channels, can be studied in the different
decay modes and the annihilation contribution measured in the
charm sector through the $D^+_s \to \pi^+\pi^-\pi^+$ decay.
\end{abstract}
\baselineskip=17pt
\section{Introduction}

The analysis of substructures should be considered in the context of 
probing hadronic-decay dynamics.
The quality of data and the availability of high-statistics samples have 
already shed a great deal of new experimental light
on the non-leptonic decays of charm particles.
Indeed, a full lifetime-hierarchy pattern has now been experimentally 
established\cite{omegac,chli}, posing severe constraints on theoretical 
models and pointing to hadronic-sector corrections.
The importance of final-state interactions (FSI), in which the two decay daughters
undergo strong rescattering after their initial formation, can be gauged 
from the isospin-amplitude interference. 
The factorization model predictions can be seriously distorted by FSI;
the example of $ D^0 \to K^+K^-/\pi^+\pi^-$ ratio,
expected to be 1.4 (BSW\cite{BSW}) and measured as $\sim$ 2.5\cite{pdg},  
underscores their important role.
In table~\ref{isospin} a list of phase shifts and isospin 
amplitudes\cite{arn} is reported: the phase shifts between 
different isospin amplitudes are seen to be often approximately $90^{\circ}$, 
indicating the general importance of FSI.

\begin{table}
\centering
\caption{ \it Isospin Decompositions.
}
\vskip 0.1 in
\begin{tabular}{|l|c|c|} \hline
 Mode  &  Ratio of amplitudes & $\delta =\delta_I -\delta_{I'}$ \\
\hline
\hline
 $K \pi$  & $ |A_{1/2}|/|A_{3/2}|    =3.99 \pm 0.25$  & 
 $ 86^{\circ} \pm 8^{\circ} $                    \\
 $K^*\pi$  & $ |A_{1/2}|/|A_{3/2}|   =5.14 \pm 0.54$  & 
 $ 101^{\circ} \pm 14^{\circ} $                    \\
 $K\rho$  & $ |A_{1/2}|/|A_{3/2}|    =3.51 \pm 0.75$  & 
 $ 0^{\circ} \pm 40^{\circ} $                    \\
 $K^*\rho$  & $ |A_{1/2}|/|A_{3/2}|  =5.13 \pm 1.97$  & 
 $ 42^{\circ} \pm 48^{\circ} $                    \\
 $K K$  & $ |A_1|/|A_0|  = 0.61 \pm 0.10$  & 
 $ 47^{\circ} \pm 10^{\circ} $                    \\
 $\pi \pi$  & $ |A_2|/|A_0|  = 0.72 \pm 0.14$  & 
 $ 82^{\circ} \pm 9^{\circ} $                    \\
\hline
\end{tabular}
\label{isospin}
\end{table}

\noindent
Recently, amplitude analysis of non-leptonic decays
has emerged as an excellent tool for studying charm hadron dynamics.
In fact, deviations from flat Dalitz plots indicate non-constant dynamics and 
a knowledge of the quantum-mechanical decay amplitudes 
allows one to properly  account for interference effects when calculating 
branching ratios. In addition, such an analysis provides new probes of FSI 
by measurement of the phase shifts between the interfering amplitudes for the 
various resonant channels. 
A problem, still unsolved in charm decay, is the reliable estimate 
of non-spectator contributions;
$D^+_s \to \pi^+\pi^-\pi^+$ is the best candidate to occur 
through an annihilation diagram since the annihilation of the two 
initial quarks is Cabibbo favoured and not suppressed as in the $D^+$.
Nevertheless, the presence of resonant channels with an $s \bar s$ 
quark content would suggest spectator processes rather than 
W annihilation, making a Dalitz-plot analysis necessary to disentangle 
direct three-body decays from resonances.
Many experiments have focused on the substructure analysis; among the 
simplest decays to analyze are those of the ground-state charm mesons into 
three pseudoscalars: \vspace{2mm}

$D\to K\pi\pi$ \qquad (Mark III\cite{markiii}, E691\cite{e691}, 
Argus\cite{Argus}, E687\cite{nostro})

$D\to K K \pi$ \qquad (E691\cite{e691_k}, E687\cite{mio})

$D \to \pi \pi \pi$ \qquad (E691\cite{e691_p}, E687 [preliminary]).

\section {\protect\boldmath{$KK\pi$} Dalitz plots}

\begin{figure}[htb]
\newsavebox{\figone}
\savebox{\figone}{\epsfig{file=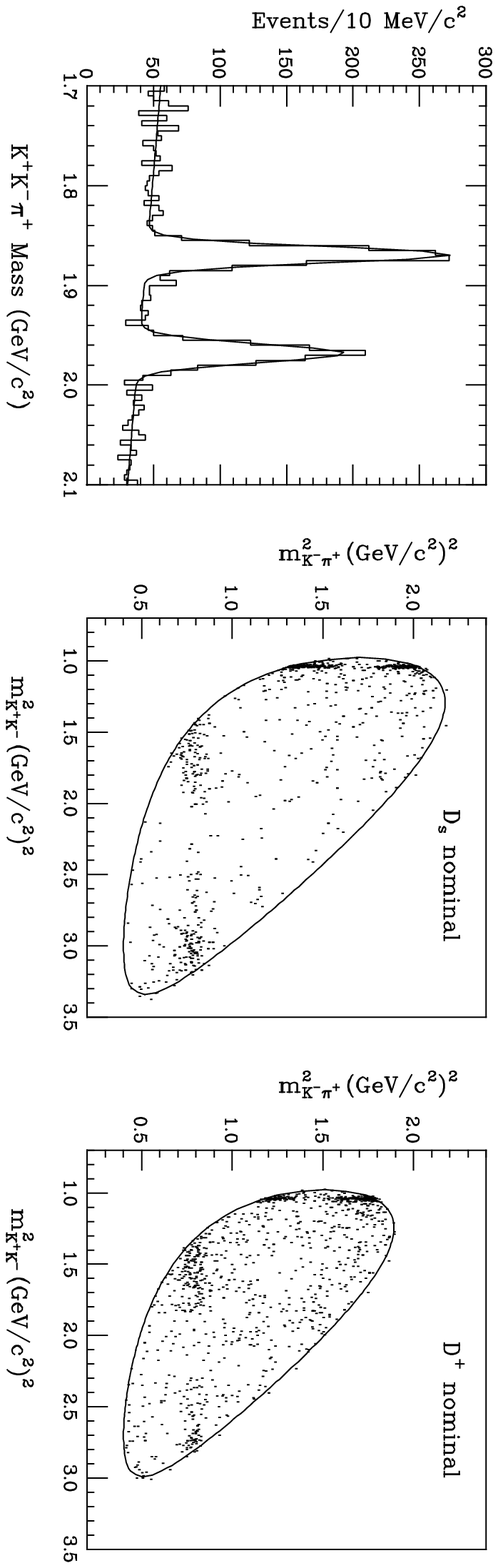,width=11.cm}}
\begin{picture}(450,150)(0,0)
 \begin{rotate}{270}
  \put(-240,-500){\usebox{\figone}}
 \end{rotate}
\end{picture}
\caption{\it $KK\pi$ mass and Dalitz plots of the $D^+$ and $D_s$ mass 
region.}
\label{figone} 
\end{figure}

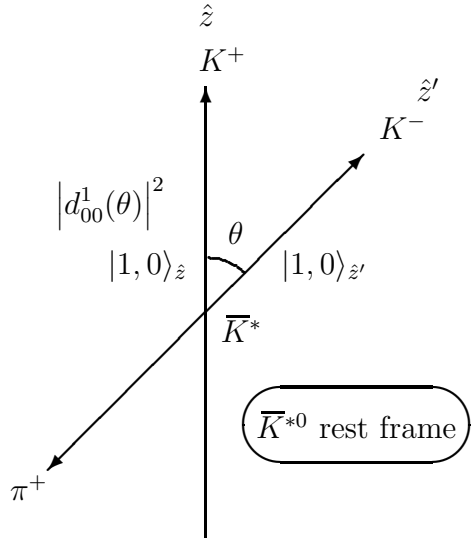
\begin{figure}[htb]
\begin{center}
\setlength{\unitlength}{1mm}
\begin{picture}(80,70)
\thicklines
\put(40,40){\vector(1,1){21}}		\put(40,40){\vector(0,1){30}}
\put(40,40){\vector(-1,-1){21}}		\put(40,40){\line(0,-1){30}}
\put(43,49){$\theta$}
\put(44.55,45.05){.}
\put(44.35,45.23){.}
\put(44.14,45.41){.}
\put(43.92,45.58){.}
\put(43.70,45.74){.}
\put(43.47,45.90){.}
\put(43.24,46.04){.}
\put(43.00,46.17){.}
\put(42.75,46.30){.}
\put(42.51,46.41){.}
\put(42.25,46.52){.}
\put(41.99,46.61){.}
\put(41.73,46.70){.}
\put(41.47,46.77){.}
\put(41.20,46.84){.}
\put(40.93,46.89){.}
\put(40.66,46.93){.}
\put(40.39,46.97){.}
\put(40.11,46.99){.}
\put(39.84,47.00){.}
\put(42,36){$\anti{K}^*$}
\put(63,63){$K^-$}	\put(68,68){$\hat{z}'$}
\put(39,72){$K^+$}	\put(39,78){$\hat{z}$}
\put(14,15){$\pi^+$}
\put(50,45){$\vert1,0\rangle_{\hat{z}'}$}
\put(27,45){$\vert1,0\rangle_{\hat{z}}$}
\put(20,53){$\displaystyle \left\vert d_{00}^1(\theta) \right\vert^2$}
\put(60,25){\oval(30,10)}
\put(45,20){\makebox(30,10){$\anti{K}^{*0}$ rest frame}}
\end{picture}
\vskip-10mm
\end{center}
\caption{\it Helicity angle definition.}
\label{helicity} 
\end{figure}

The $D^+_s, D^+ \to K^+ K^-\pi^+$ \footnote{Through this paper the 
charge conjugate state is always implied.}
Dalitz plots in $m^2_{KK},m^2_{K\pi}$, obtained by E687 and shown 
in fig.~\ref{figone}, are particularly instructive.
The $D^+_s$ Dalitz is very highly dominated by the $\phi \pi^+$ and 
$\anti{K}^{*0} K^+ $ channels while the $D^+$ has a significant additional 
contribution appearing as a uniform event intensity.  
Depopulation of the central region of the $\phi$ and 
$\anti{K}^{*0}$ is due to a node in the decay function, reflecting 
angular-momentum conservation; the $\cos{\theta_{helicity}}$ 
(fig.~\ref{helicity}) dependence is due to the vector nature of the 
mesons ($\phi, \anti{K}^{*0}$) decaying into two pseudoscalars. 

\begin{figure}[htb]
\begin{center}
\setlength{\unitlength}{1mm}
\begin{picture}(100,45)(0,0)
\thicklines
\put(30,25){\circle{8}} \put(27,24){$F_D$} 
\put(70,25){\circle{8}} \put(68,24){$F_r$} 
\put(05,24){$(\pi+D)_\mu$} 
\put(78,24){$(K^+-K^-)_\nu$} 
\put(48,28){$\phi$}
\put(25,03){\makebox(50,20)	{$\frac
					{g^{\mu\nu}-q^\mu q^\nu/m_0^2}
					{q^2-(m_0-i\Gamma/2)^2}		$}}
\put(04,47){$\pi$}	\put(93,47){$K^+$} 
\put(04,00){$D$}	\put(93,00){$K^-$} 
\put(27,28){\vector(-1,1){10}}		\put(73,28){\vector(1,1){10}}
\put(18,37){\line(-1,1){10}}		\put(82,37){\line(1,1){10}}
\put(27,22){\line(-1,-1){10}}		\put(73,22){\line(1,-1){10}}
\put(08,03){\vector(1,1){10}}		\put(92,03){\vector(-1,1){10}}
\multiput(35,25)(4.8,0){7}{\line(1,0){2.0}}
\end{picture}
\vskip -.5 cm
\end{center}
\caption{\it $D \to KK \pi$ decay diagram.}
\label{model} 
\end{figure}
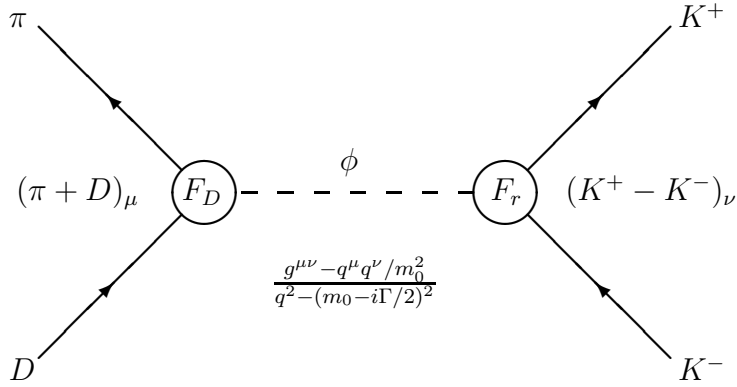

\noindent
The Dalitz plots are fit with a coherent sum of 
resonances: quasi-two-body channels of the type 
$$
\begin{array}{rcl} 
 D & \to & r + c     \\
   &     & \spig a + b,
\end{array} 
$$
which are described by a decay function (eq.~\ref{eq1} and fig.~\ref{model}), 
\begin{equation}
{\cal M} = F_D \,F_r \times |
\bar c|^J |\bar a|^J\,P_j(\cos{\theta_{ac}^r)}
\times BW(m_{ab}). 
\label{eq1}
\end{equation}
The matrix element is the product of 
\begin{itemize}
\item  two vertex form factors (Blatt-Weisskopf momentum-dependent factors),
\item an order $J$ Legendre polynomial representing the decay angular 
wave function,
\item a relativistic Breit-Wigner (BW).
\end{itemize}

\begin{figure}[htb]
\newsavebox{\asymme}
\savebox{\asymme}{\epsfig{file=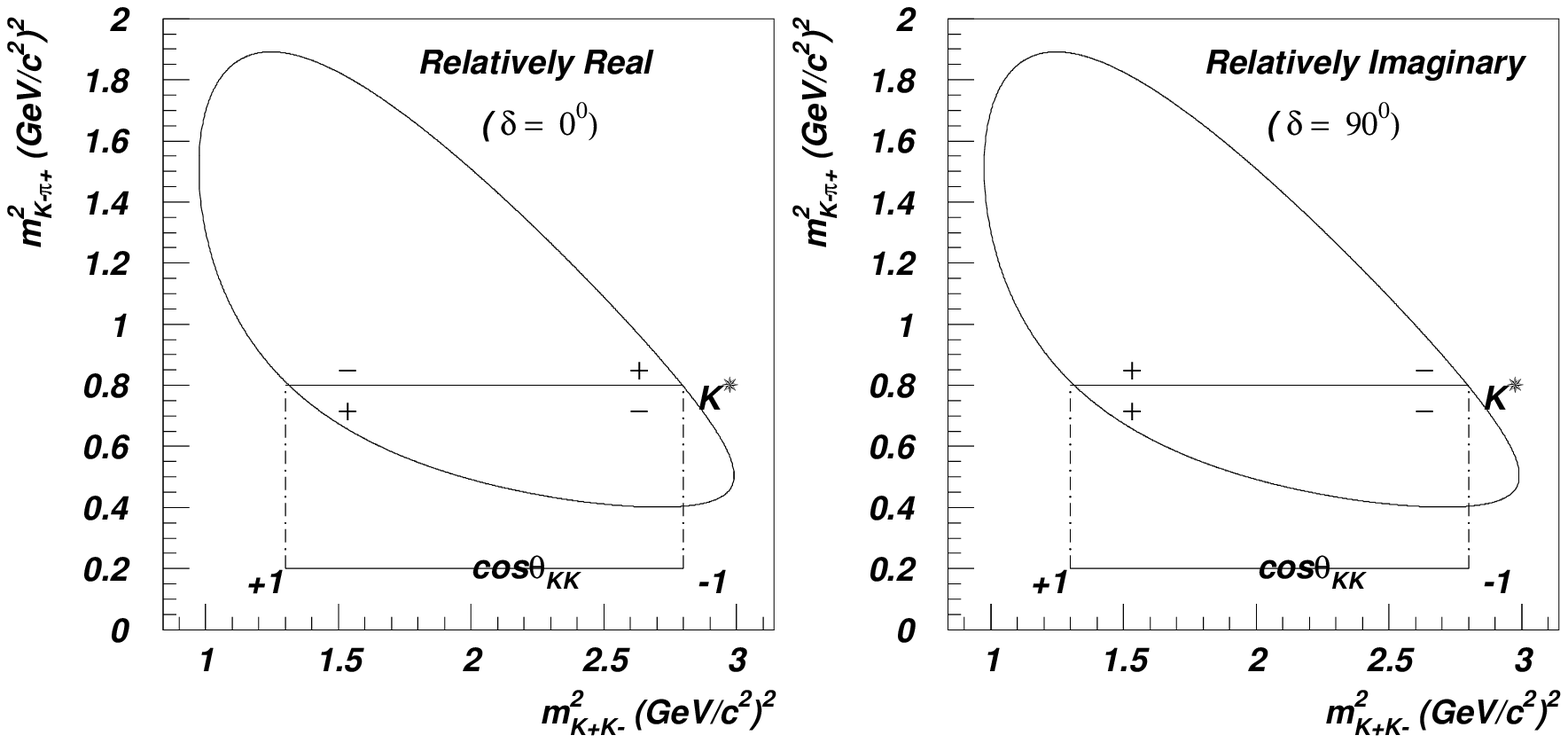,width=15.cm}}
\begin{picture}(200,170)(0,0) 
  \put(10,-220){\usebox{\asymme}}
\end{picture}
\caption{\it The interference pattern of $D^+\to KK\pi$ in the 
$\protect\anti{K}^{*0}$ band.}
\label{asymme} 
\end{figure}

\noindent
The total decay amplitude consists of various functions of the form 
(\ref{eq1}), each multiplied by a complex factor 
$a_r e^{i\delta_r}$; in the simpler case of the $D^+_s$, where 
only two contributions dominate ($\phi \pi^+$ and $\anti{K}^{*0}K^+$)
the total amplitude is 
\begin{equation}
\begin{array}{cl}
{\cal A} (D^+_s \to K^+K^-\pi^+) & =   a_{\anti{K}^{*0}}
e^{i\delta_{\anti{K}^{*0}}} 
{\cal M} (\pi^+ K^- K^+|\anti{K}(892)^{*0}) \\ 
& \\
& \null +  a_{\phi}e^{i\delta_{\phi}}
{\cal M}(K^+ K^-\pi^+|\phi),
\end{array}
\label{eq2}
\end{equation}
where the modulus $ a$ measures the importance of the channel in the 
decay and $\delta$ is the phase.
With the factorization hypothesis, the bare amplitudes 
will be real. Phases are then induced by FSI as consequence of 
Watson's theorem, which states that the observed amplitudes are related 
to the bare amplitudes by the square root of a strong-interaction S-matrix, 
describing the hadron-hadron scattering, 
\begin{equation}
\left( \begin{array}{c} a_{\anti{K}^{*0}}e^{i\delta_{\anti{K}^{*0}}} \\[1mm] 
a_{\phi}e^{i\delta_{\phi}}
 \end{array} \right)_{\! \! \! obs} =
\left( \begin{array}{cc} \eta e^{2i\delta_1} & 
i\sqrt{1-\eta^2} e^{i(\delta_1+\delta_2)} \\[1mm] 
i\sqrt{1-\eta^2}e^{i(\delta_1+\delta_2)} & \eta e^{2i\delta_2} 
\end{array} \right)^{1/2}   
\left( \begin{array}{c} a_{\anti{K}^{*0}} \\[1mm]  
a_{\phi} \end{array} \right)_{\! \! \! bare} 
\label{mat1}
\end{equation}
through the strong-interaction phase shifts ($\delta_{1,2}$)
and the elasticity parameter ($\eta$).
While FSI can be measured in two-body decays through the interference of 
isospin amplitudes, in three-body decays they can be inferred from the 
interference of resonant channels, i.e., from the phase shifts.
In practice, the phase of one resonance channel is factored out and hence 
relative phases are measured from experimental fits to ${\cal A^* A }$.
Visual evidence for FSI effects is possible through the study of 
the behaviour of the 
$\anti{K}^{*0}$ lobes in $D^+ \to K^+K^-\pi^+$ (fig.~\ref{figone}). 
The asymmetry, not present in the $D^+_s$ case, can be 
interpreted as an interference effect (fig.~\ref{asymme}) 
with a nearly constant amplitude (a broad scalar), which may be written 
as $\cos{\delta} +i \,\,\sin{\delta}$.
The interference of these two amplitudes will produce a contribution
to the intensity of the form 
\begin{eqnarray}
2 Re\{ (\cos\delta +i \,\sin\delta)^* \frac{\cos\theta_{KK}} 
{m^2_r-m^2_{K\pi} - i\Gamma m_r} \} &&  \nonumber \\ 
=\; 2\frac{(m^2_r-m^2_{K\pi}) \cos\theta_{KK}\cos\delta}
{(m^2_r-m^2_{K\pi})^2 +\Gamma^2m^2_r} &+& 
2 \frac{\Gamma M_r \cos{\theta_{KK}} \sin\delta}
{(m^2_r-m^2_{K\pi})^2 +\Gamma^2m^2_r}. 
\label{eq3}
\end{eqnarray}

\noindent
As one moves along the $\anti{K}^{*0}$ band in the direction of 
increasing $m^2_{KK}$, $\cos\theta$ changes from +1 to $-$1 and 
both terms in eq.~\ref{eq3} switch sign.
Along a line of constant $m^2_{KK}$ the first term, which dominates for a real
amplitude ($\delta\sim 0^{\circ}$), switches sign on passing through 
the resonance pole (thus cancelling the interference), while the second 
interference term, which dominates for a relatively imaginary amplitude 
does not change sign under this motion.
Thus, the patterns for real and imaginary phases are quite different; the 
interference present in the $D^+$ more closely resembles the pattern for 
a relatively imaginary amplitude, $\delta \sim 90^{\circ}$. 
A fully coherent Dalitz analysis of $D^+\to K^+K^-\pi^+$ confirms this
supposition (see table~\ref{comkkp}). 

\subsection{ Speculation on the universality of Dalitz phases }

Since the phase factors in the $D^+$ and $D^+_s$ both originate from FSI, 
one is tempted to speculate that they might be the same.
In fact, if the $D^+$ and $D^+_s$ were degenerate in mass, 
the $S$-matrix describing 
the rescattering for $\phi \pi^+$ and $\anti{K}^{*0} K$ would be the same:
\begin{equation}
S = \left( \begin{array}{cc} \eta e^{2i\delta_1} & 
i\sqrt{1-\eta^2} e^{i(\delta_1+\delta_2)} \\[1mm]
i\sqrt{1-\eta^2}e^{i(\delta_1+\delta_2)} & \eta e^{2i\delta_2} 
\end{array} \right).
\label{mat2}
\end{equation}
For elastic FSI ($\eta =1$) it is easy to see that the observed 
amplitudes would then
pick up a relative phase shift given by $\delta_1-\delta_2$, which is 
common to $D^+_s$ and $D^+$ in the limit of degenerate masses.
In this case, defining $R = (a_{\phi} / a_{\anti{K}^{*0}} )_{phys.}$ and 
$R_0 = (a_{\phi} / a_{\anti{K}^{*0}})_{bare}$,
\begin{equation}
\left( \begin{array}{c}  R \\[1mm] 
1 \end{array} \right) =
\left( \begin{array}{cc} e^{2i\alpha_{\phi}} & 0 \\[1mm]
0 & e^{2i \alpha_{K^*}} \end{array} \right)^{1/2}
\left( \begin{array}{c}  R_0 \\[1mm] 
1 
\end{array} \right) 
\label{eq4}
\end{equation}
and thus $\delta_{\phi}-\delta_{\anti{K}^{*0}} = 
\alpha_{\phi} -\alpha_{\anti{K}^{*0}}$ would be equal 
for both $D^+$ and $D^+_s$.
If $\eta <1$, the observed phases would depend on the ratio of the $\phi$ and
$\anti{K}^{*0}$ amplitudes, which are different in the $D^+_s$ and $D^+$
and thus $\delta_{\phi}-\delta_{\anti{K}^{*0}}(D^+)  \ne 
\delta_{\phi}-\delta_{\anti{K}^{*0}}(D^+_s)$.
In the particular case of total 
inelasticity, $\eta=0$, 
\begin{equation}
\left( \begin{array}{c}  R \\[1mm] 
1 \end{array} \right) =
\left( \begin{array}{cc}  0 & i \\[1mm]
i & 0 \end{array} \right)^{1/2}
\left( \begin{array}{c}  R_0 \\[1mm] 
1 \end{array} \right) = \frac{1}{\sqrt 2}
\left( \begin{array}{cc}  1 & i \\[1mm]
i & 1 \end{array} \right)
\left( \begin{array}{c}  R_0 \\[1mm]
1 \end{array} \right) 
\label{eq5}
\end{equation}
and $\delta_{\phi} - \delta_{K^*} = 
90^{\circ} \,- \,2 \,\tan^{-1} R_0$.

\noindent
Thus, the universality of Dalitz amplitude phases might provide a clue to the
elasticity of FSI. 
Table~\ref{comkkp} compares the E687 final results for $D^+$ and
$D^+_s$\cite{mio} obtained through a fully coherent analysis, the total fit
amplitude containing various $K^+K^-$ and $K^-\pi^+$ resonances plus a flat
non-resonant term. 
\begin{table}
\centering
\caption{\it Comparison of $D^+$ and $D^+_s \to 
K^+K^-\pi^+$.\protect\footnotemark}
\vskip 0.1 in
\begin{tabular}{|c|c|c|}\hline
Parameter & $D^+$     & $D^+_s$ \cr\hline
\hline 
$\delta_{\anti{K}(892)^{*0} K }$ & $0^{\circ}$ (fixed) & $0^{\circ}$(fixed) 
$\anti{K}^*$\cr\hline
$\delta_{\phi\pi}$ & -159 $\pm$ 8 $\pm$ 11$^{\circ}$ & 178 $\pm$ 20 $\pm$ 
24$^{\circ}$ 
\cr\hline
$\delta_{\anti{K}(1430)^{*0} K }$ & 70 $\pm$ 7 $\pm$ 4$^{\circ}$ & 
152 $\pm$ 40 $\pm$ 39$^{\circ}$ \cr\hline 
$\delta_{f_0 (980) \pi}$    & - & 159 $\pm$ 22 $\pm$ 16$^{\circ}$  \cr\hline
$\delta_{f_{j(1710)}\pi}$   & - & 110 $\pm$ 20 $\pm$ 17$^{\circ}$  \cr\hline
\hline
$f_{\anti{K}(892)^{*0} K} $  & 0.301 $\pm$ 0.020 $\pm$ 0.025 & 
0.478 $\pm$ 0.046 $\pm$ 0.040  \cr\hline
$f_{\phi \pi}$	             & 0.292 $\pm$ 0.031 $\pm$ 0.030 & 
0.396 $\pm$ 0.033 $\pm$ 0.047   \cr\hline
$f_{\anti{K}(1430)^{*0} K}$  & 0.370 $\pm$ 0.035 $\pm$ 0.018 & 
0.093 $\pm$ 0.032 $\pm$ 0.032   \cr\hline
$f_{f_0(980)  \pi}$	     & - & 0.110 $\pm$ 0.035 $\pm$ 0.026   \cr\hline
$f_{f_{j(1710)}  \pi}$       & - & 0.034 $\pm$ 0.023 $\pm$ 0.035   \cr\hline
\end{tabular}
\label{comkkp}
\end{table}
A few comments may be made on these results.
The $D^+$ consists of nearly equal contributions of $\anti{K}(892)^{*0} K^+,
\phi \pi$ and the broad scalar $\anti{K}(1430)^{*0} K$ while the 
$D^+_s$ is strongly dominated by just $\anti{K}^{*0} K^+$ and $\phi \pi^+$ plus
a small contribution from $f_0(980)$ and $f_j(1710)$, which sum to 14\%.
Both the $D^+$ and $D^+_s$ can be fit by entirely quasi-two-body
processes without the inclusion of a non-resonant contribution.
It is interesting to note that both charm states have relatively 
real phases between the dominant $\anti{K}^{*0} K^+$ and $\phi \pi^+$ channels,
$\sin{(\delta_{\phi}-\delta_{\anti{K}^{*0}})} \approx  0$, which indicate 
absent or cancelling FSI phase shifts.
In contrast, for the $D^+$, $\sin{(\delta_{\anti{K}(1430)^{*0}}-
\delta_{\anti{K}(892)^{*0}})} \approx 0.94$, pointing to strong 
FSI effects, as anticipated in the discussion of the asymmetric 
$\anti{K}^{*0}$ lobes, 
with the $\anti{K}(1430)^{*0}$ playing the role of the nearly constant scalar 
resonance.
Unfortunately, the large error, 
$\sigma(\Delta\delta) = 56^{\circ}$,
on the $\anti{K}(1430)^{*0} K$ relative phase
for the $D^+_s$ precludes a meaningful test of the conjecture of a 
universal phase shift for this channel.

\section{E687 preliminary results on 
\protect\boldmath{$D^+_s,D^+\to\pi^+\pi^-\pi^+$}}

\begin{figure}[htb]
\newsavebox{\ppp}
\savebox{\ppp}{\epsfig{file=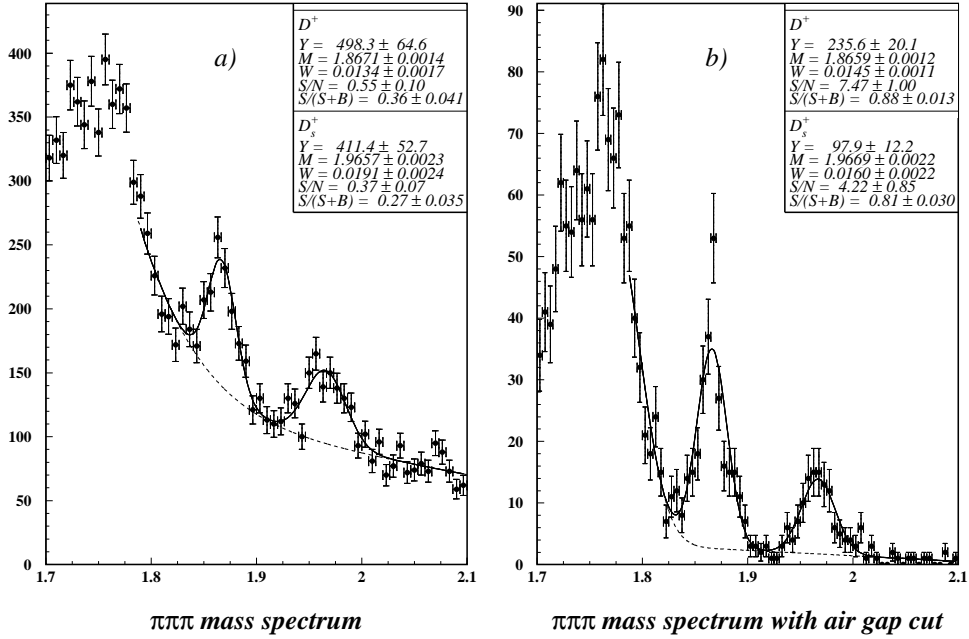,width=14.cm}}
\begin{picture}(460.,250.)(0,0)
  \put(20,0){\usebox{\ppp}}
\end{picture}
\caption{\it $\pi\pi\pi$ mass spectrum a) without and b) with the air-gap cut.
}
\label{ppp} 
\end{figure}

\begin{figure}[htb]
\newsavebox{\folded}
\savebox{\folded}{\epsfig{file=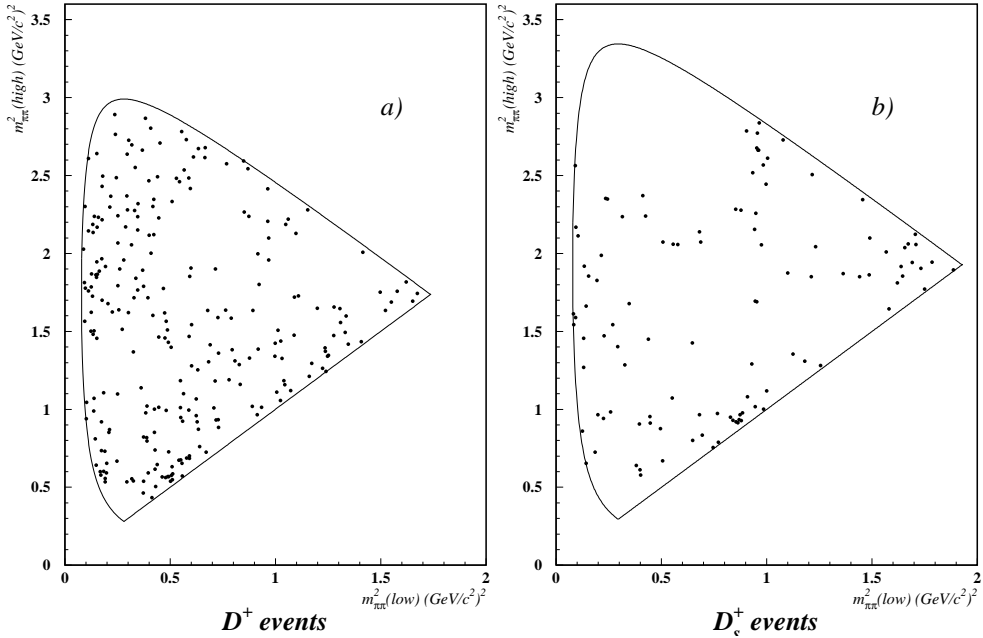,width=14.cm}}
\begin{picture}(460.,250.)(0,0) 
  \put(20,0){\usebox{\folded}}
\end{picture}
\caption{\it Folded Dalitz plots for a) $D^+$and b) $D^+_s \to \pi\pi\pi$.
}
\label{folded} 
\end{figure}

\footnotetext{The fractions $f_r$ are known as decay fractions and represent
the ratio of the integrated Dalitz intensity for a single resonance $r$ divided
by the intensity with all contributions present 
$$
\mbox{decay fraction}_i = \frac{\int dm^2_{ab}dm^2_{ac}( a_i e^{i\delta_i}
{\cal M } (abc |r))^*(a_ie^{i\delta_i}{\cal M }(abc|r))}
{\int dm^2_{ab} dm^2_{ac} (\sum_i a_ie^{i\delta_i}{\cal M} (abc|r))^*
(\sum_i a_i e^{i\delta_i} {\cal M}(abc|r))}. 
$$}
The observation of a non-resonant three-pion decay of the $D^+_s$ would
represent a clear signature in favour of non-spectator processes in the charm
sector. 
To determine the rate of $D^+_s\to\pi^+\pi^-\pi^+$ due to annihilation, the
contributions from the resonant decays that most likely originate from the 
$s\bar s$ quark through spectator-diagrams have to be removed. 
The power of the E687 analysis is the {\it air-gap cut}, which imposes the 
requirement that the secondary vertex lies in the air-gap region downstream 
of the $Be$ interaction target and upstream of any spectrometer elements.
It significantly reduces the non-charm background and increases the 
signal-to-noise ratio by about a factor of four in both the $D^+$ and 
$D^+_s$ cases.
In fig.~\ref{ppp} the invariant mass plots are shown with and without 
the air-gap cut. 
The large $K \pi\pi$ contamination, with the $K$ misidentified as a $\pi$, is 
removed by requiring the candidate mass to be incompatible with the 
$D^+\to K\pi\pi$ hypothesis.
The Dalitz plots for the $D^+$ and $D^+_s$ signal regions, defined 
within $\pm 2\sigma$ from the mass peak, are shown in 
fig.~\ref{folded}. 
Owing to the presence of two identical particles, the $\pi^+\pi^-$ amplitudes 
have to be Bose symmetrized; the representation chosen is that 
of a {\it folded} Dalitz plot where the lower and the higher value of the 
two possible $m^2_{\pi^+\pi^-}$ combinations are plotted on the abscissa and 
ordinate respectively.

\begin{figure}[htb]
\newsavebox{\proj}
\savebox{\proj}{\epsfig{file=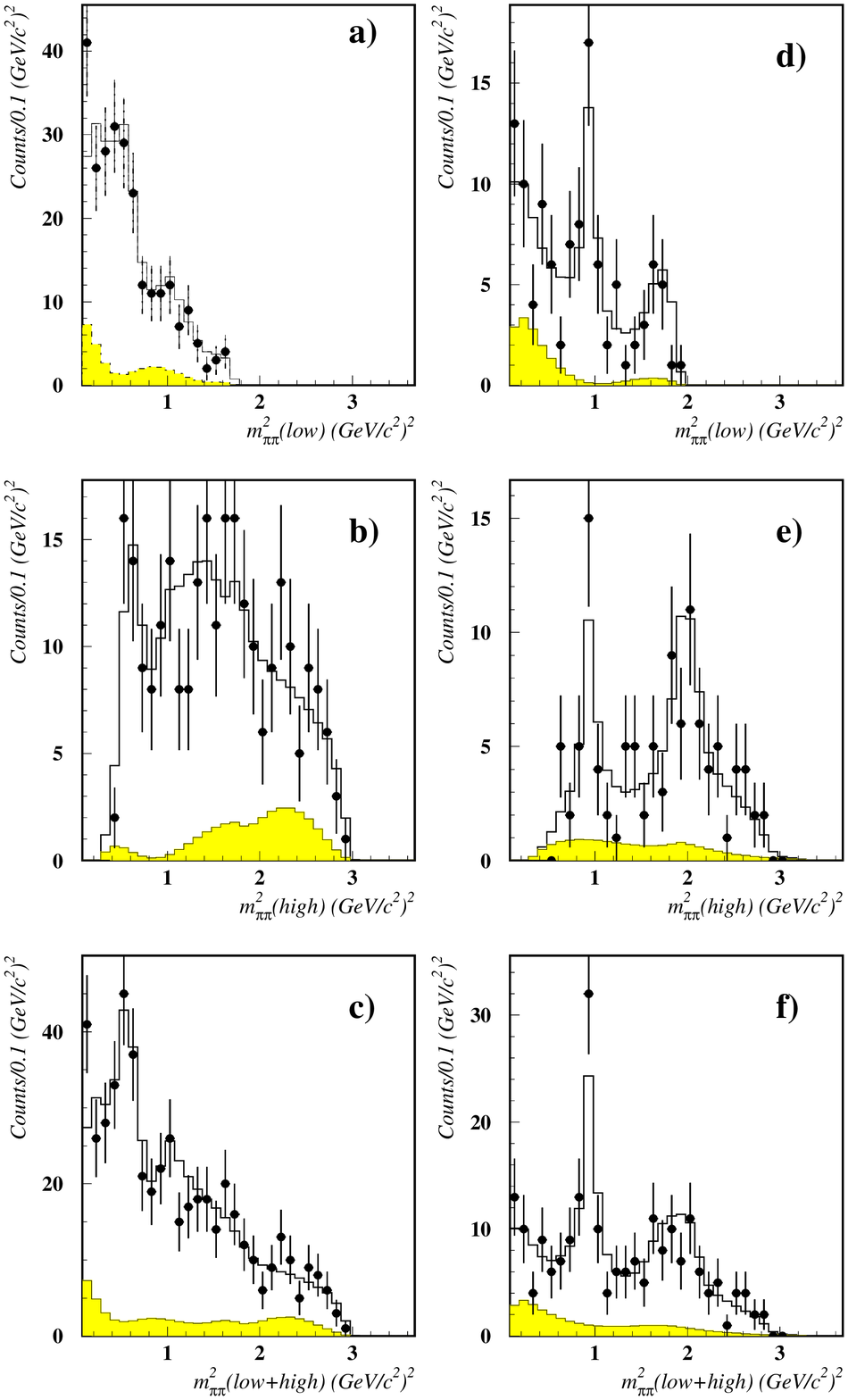,width=14.cm}}
\begin{picture}(460.,600.)(0,0) 
  \put(20,10){\usebox{\proj}}
\end{picture}
\caption{\it Invariant $\pi^+\pi^-$ mass projections for the $D^+$:
a) low, b) high, c) sum; and $D^+_s$: 
d) low, e) high, f) sum. 
The crosses are data, the continuous line is the fit 
result and the shaded area represents the background.
}
\label{proj} 
\end{figure}

\noindent
Following the model already explained, the total amplitude is assumed
to consist of a flat uniform term representing the non-resonant
contribution plus a sum of functions representing the intermediate strong
resonances and decay angular-momentum conservation. The fit parameters 
are then the amplitude coefficients $a_i$ and phases $\delta_i$.
All the resonances decaying into $\pi^+\pi^-$ with sizeable branching 
fractions have been considered in the fit function.
Fig.~\ref{proj} shows the two mass projections of the Dalitz plots 
together with their sum.
The resulting fit fractions and the relative phases are listed in 
table~\ref{comp3p}. 

\begin{table}
\centering
\caption{\it $D^+$ and $D^+_s \to \pi^+\pi^-\pi^+$ comparison.}
\vskip 0.1 in
\begin{tabular}{|c|c|c|}\hline
Parameter & $D^+$     	   & $D^+_s$ \cr\hline\hline
$\delta_{nr}$ 		   & $0^{\circ}$ (fixed)    & 235 $\pm$ 22$^{\circ}$  
\cr\hline
$\delta_{\rho\pi}$         & 27 $\pm$ 14$^{\circ}$  & 53 $\pm$ 44$^{\circ}$     
\cr\hline 
$\delta_{f_2(1270) \pi}$   & 207 $\pm$ 17$^{\circ}$ & 100 $\pm$ 18$^{\circ}$
\cr\hline
$\delta_{f_0(980) \pi}$    & 197 $\pm$ 28$^{\circ}$ & 0$^{\circ}$ (fixed)     
\cr\hline
$\delta_{f_{\cal{S}}(1500)\pi}$ & - & 234 $\pm$ 15$^{\circ}$    
\cr\hline\hline
$f_{nr}$	      	   & 0.589 $\pm$ 0.105    & 0.121 $\pm$ 0.115
\cr\hline
$f_{\rho \pi}$	           & 0.289 $\pm$ 0.055    & 0.023 $\pm$ 0.027       
\cr\hline
$f_{f_2 (1270)  \pi}$	   & 0.052 $\pm$ 0.034    & 0.123 $\pm$ 0.056       
\cr\hline
$f_{f_0 (980)  \pi}$	   & 0.027 $\pm$ 0.031    & 1.074 $\pm$ 0.140 
\cr\hline
$f_{f_{\cal{S}}(1500)\pi}$    & - & 0.274 $\pm$ 0.114      
\cr\hline
\end{tabular}
\label{comp3p}
\end{table}

\noindent
The $D^+$ appears to be largely dominated by the non-resonant and 
$\rho \pi^+$, with marginal contributions from the scalar $f_0(980)$ 
and the spin-2 resonance $f_2(1270)$.
A signature for the $\rho(770)$ is the peak in the mass projections
around 0.6 (GeV/c$^2)^2$ and for the $f_0(980)$ the bump around 
1.0 (GeV/c$^2)^2$. 
The $D^+_s$ Dalitz plot, in contrast to the $D^+$, is more structured.
Even a visual inspection reveals the $f_0(980)$ band, previously observed 
by E691\cite{e691_p}, and an accumulation
of events at the sharp corner on the right side of the Dalitz plot, which
could be attributed to the third helicity lobe of the $f_2(1270)$ plus an
accumulation near the contour on the left side corresponding to the first 
helicity-lobe region of the $f_2(1270)$.  
The non-resonant contribution appears to be marginal because of the 
absence of population in the upper corner of the Dalitz plot.

\noindent
By way of contrast, a uniform-density horizontal band around 
2.0 (GeV/c$^2)^2$ is manifest, whose structure suggests 
the presence of a scalar resonance in $\pi^+\pi^-$ mass around 
1.5 (GeV/c$^2)$.
Given the major uncertainty surrounding dipion scalar resonances in this mass region,
the experiment decided to use its own data to find the parameters for 
a single state, which best reproduces the Dalitz plot.
Several fits of $D_s^+$ Dalitz plot were performed by
varying the mass (1200 MeV -- 1600 MeV) and the width (25 MeV -- 400 MeV) of 
this state, which was parametrized as a Breit-Wigner resonance.
Fits were performed taking into account all possible additional 
contributions from
a non-resonant component and the following well-established resonances, 
$f_0(980)$, $f_2(1270)$ and $\rho(770)$. 
A clear maximum of the likelihood function is found at $m=1.475$~GeV/c$^2$ and 
$\Gamma=100$~MeV/c$^2$. 
The parameters of this state, which is denoted as ${\cal{S}}(1500)$, 
are remarkably consistent with the $f_0(1500)$ entry of 
PDG96.\footnote{Although the mass and width of this state are in 
excellent agreement with the $f_0(1500)$ entry in PDG96, it is to be 
emphasized that several interfering resonances could equally well describe 
the $D_s^+$ Dalitz plot.}
Evidence for the non-resonant channel, as well as the $\rho \pi^+$ 
contribution, would have important implications for the theory;
while the latter decay is expected to be heavily suppressed in the 
factorization model assuming isospin symmetry\cite{mauro}, 
the evidence of the non-resonant channel would indicate the presence of 
an initial-quark annihilation process.
However, in contrast with the $\rho(770)\pi^+$ case, 
assessing the presence of the non-resonant component in a decay is a very 
difficult task, which would at least require a high-statistics sample since 
a coherent sum of wide resonances could easily mimic an almost flat 
contribution. 
A comparison of the $D^+$ and $D^+_s \to \pi^+\pi^-\pi^+$ phase shifts 
(table~\ref{comp3p}) suggests a different and presumably more complex
dynamics with respect to the $K^+K^-\pi^+$, with a different role of the 
FSI for $D^+$ and $D^+_s$ in this decay channel.

\section{\protect\boldmath{$K\pi\pi$} Dalitz plots}

The $K \pi\pi$ final state is the most studied so far: many experiments 
have focused on it and results are published for 
$D^+\to K^-\pi^+\pi^+$, $D^0\to K^0_s \pi^+\pi^-$ and
$D^0\to K^-\pi^+\pi^0$.  
A certainly striking conclusion reached by all groups is 
the presence of a large non-resonant 
component for $D^+ \to K^- \pi^+ \pi^+$, which is unique to this decay, 
the majority of charm-meson decays being dominated by quasi-two-body
processes.
Furthermore, amplitude interference is significant in all three 
modes; the sum of the decay fractions, reported in table~\ref{penul}, 
greatly exceeds unity\cite{nostro}.

\begin{table}
\centering
\caption{\it Sum of the $K\pi\pi$ final state decay fractions.}
\vskip 0.1 in
\begin{tabular}{|c|c|c|c|} \hline 
& $K^- \pi^+ \pi^+$ & $\anti{K}^0 \pi^+\pi^-$ & $K^- \pi^+\pi^0$ \cr \hline
\hline 
$\sum f_r$ & 1.47 & 1.27 & 1.18 \cr\hline
\end{tabular}
\label{penul}
\end{table}

\noindent
An analysis of the phase shifts is interesting also in this case:
in table~\ref{ultima} the E687 $|\sin{(\delta - \delta_{\anti{K}^* \pi})}|$
is reported for the three modes. 

\begin{table}
\centering
\caption{ \it $|\sin{(\delta - \delta_{\protect\anti{K}^* \pi})}|$ for 
the three $D^+\to K \pi \pi$ modes.
}
\vskip 0.1 in
\begin{tabular}{|c|c|c|c|} \hline 
& $K^- \pi^+ \pi^+$ & $\anti{K}^0 \pi^+\pi^-$ & $K^- \pi^+\pi^0$ \cr \hline
\hline
$\anti{K} \rho$ & - & 0.69 $\pm $ 0.1 & 0.31 $\pm$ 0.2 \cr\hline
$\anti{K}^*(1430) \pi$ & 0.26 $\pm$ 0.09 & 0.24 $\pm $ 0.23 & - \cr\hline
nr & 0.74 $\pm $ 0.03  & - & 0.97 $\pm$ 0.25 \cr\hline
\end{tabular}
\label{ultima}
\end{table}

\noindent
It is interesting to note that the $\sin{\Delta \delta}$ values are 
consistent row by row: inelastic FSI or isospin phase shift could break this. 
One might also be tempted to affirm the presence of large FSI 
for $\anti{K}^* \pi$ relative to the non-resonant channel, but the issue
here is to understand the precise nature of the non-resonant contribution: i.e., 
whether it is a single process or a mixture of broad resonances.
In addition, although the amplitude model qualitatively  reproduces 
many features of the data, statistically significant discrepancies 
are observed in some of the fits.
This may be suggestive of strong-interaction dynamics not contained in the 
model, such as the presence of new, undiscovered wide resonances or 
a non-uniform, non-resonant amplitude with a possibly varying phase.

\section{Conclusions}

Dalitz-plot analysis has emerged as a powerful tool for studying 
hadronic charm decay, providing a new handle in dealing with FSI.
In this context the recent high-statistics data have already furnished 
new interesting results: fully coherent analyses are now available 
for the $KK\pi$, $\pi\pi\pi$ and $K \pi\pi$ final states of the 
$D$ mesons. 
These decays are found to be dominated by two-body resonances with 
the unique exception of $D^+ \to K^-\pi^+\pi^+$.  
Comprehension of the whole decay pattern depends on 
a correct parametrization of the non-resonant contribution, which 
could influence evaluation of amplitudes and phases. 
Attempts to improve the non-resonant flat dynamics, allowing a variation over 
the Dalitz plot, have been proposed\cite{Bediaga} and will be implemented in 
the next analyses.
Looking to the future prospects: exciting data are expected from the 
new E831 run (now in progress) while E791 should give results pretty soon.
When the quality of the data allows a study of CP violation in the 
charm-meson sector through the measurement of the $D$ and 
$\anti{D}$ phases, a reliable estimate will require the ability to  
factor out the FSI induced phases.

\section{Acknowledgements}
I wish to acknowledge Prof. J.E.~Wiss as being the first to address 
important issues in the Dalitz analysis and Drs. D.~Menasce, 
L.~Moroni, D.~Pedrini and P.G.~Ratcliffe for useful discussions.


\begin{thebibliography}{99}
\bibitem{omegac} P.L.~Frabetti { \it et al.},
Phys. Lett. {\bf B357}, 678 (1995).
\bibitem{chli} 
S.~Malvezzi, {\em Charm lifetime}: to appear in the 
Proc. Sixth International Symposium on Heavy Flavours Physics, Pisa,
June 1995, 
e-print: hep-ph/9507391.
\bibitem{BSW} M.~Bauer, B.~Stech and M.~Wirbel, 
Z.~Phys. {\bf C34}, 103 (1987).
\bibitem{pdg} PDG, R.M.~Barnett {\it et al.},
Phys. Rev. {\bf D54} (1996).
\bibitem{arn} T.~Browder, K.~Honsheid and D.~Pedrini,
to appear in Ann. Rev. Nucl. Part. Sci. (1996).
\bibitem{markiii} J.~Adler {\it et al.},
Phys. Lett. {\bf B196}, 107 (1987).
\bibitem{e691} J.C.~Anjos {\it et al.},
Phys. Rev. {\bf D48}, 56 (1993).
\bibitem{Argus} H.~Albrecht {\it et al.},
Phys. Lett. {\bf B308}, 435 (1993).
\bibitem{nostro} P.L.~Frabetti {\it et al.},
Phys. Lett. {\bf B331}, 217 (1994).
\bibitem{e691_k} J.C.~Anjos {\it et al.},
Phys. Rev. Lett. {\bf 60}, 897 (1988).
\bibitem{mio} P.L.~Frabetti {\it et al.},
Phys. Lett. {\bf B351}, 591 (1995).
\bibitem{e691_p} J.C.~Anjos {\it et al.},
Phys. Rev. Lett. {\bf 62}, 125 (1989).
\bibitem{Bediaga} I.~Bediaga, C.~Gob\'{e}l and R.~Mendez-Galain,
CBPF-NF-029 (May 1996), \\
e-print: hep-ph/9605442. 
\bibitem{mauro}M.~Anselmino, I.~Bediaga and E.~Pedrazzi,
CBPF-NF-045 (July 1995), \\
e-print: hep-ph/9507292.

\end{thebibliography}
\end{document}